\documentclass[aps,prd,preprint,preprintnumbers,groupaddress,eqsecnum,showpacs]{revtex4-1}
\usepackage[T1]{fontenc}
\usepackage[utf8]{inputenc}
\usepackage{latexsym}
\usepackage{amsmath}
\usepackage{amsfonts}
\usepackage{graphicx, color}
\input{definitions.command}
\usepackage[colorlinks=false]{hyperref}
\usepackage[all]{hypcap}
\usepackage{pgfplotstable}
\usepackage{booktabs}
\usepackage{array}
\usepackage{colortbl}
\usepackage{natbib}
\usepackage{cancel}
\usepackage[caption=false]{subfig}
\pgfplotstableset{/pgfplots/table/string type , precision=3}

\usepackage{ulem}

\newcommand{\Sgn}{{\rm sgn}}
\newcommand{\Htilde}{\tilde H_m^{-1}}

\newcommand{\Htildesq}{\tilde H_m^{-2}}
\newcommand{\Dtildesq}{\tilde D_m^{-2}}

\begin{document}

\preprint{KEK-CP-326, OU-HET-871}

\flushbottom

\title{Violation of chirality of the M\"obius domain-wall Dirac operator from the eigenmodes}
\author{Guido Cossu}
\email[]{cossu@post.kek.jp}
\affiliation{Theory Center, IPNS, High Energy Accelerator Research Organization (KEK),\\
  Tsukuba, Ibaraki 305-0810, Japan}

\author{Hidenori Fukaya}
\email[]{hfukaya@het.phys.sci.osaka-u.ac.jp}
\affiliation{Department of Physics, Graduate School of Science, Osaka University,
  Toyonaka, Osaka 560-0043, Japan}

\author{Shoji Hashimoto}
\email[]{shoji.hashimoto@kek.jp}
\affiliation{Theory Center, IPNS, High Energy Accelerator Research Organization (KEK),\\
  Tsukuba, Ibaraki 305-0810, Japan}
\affiliation{School of High Energy Accelerator Science, The Graduate University for Advanced Studies (Sokendai), Tsukuba 305-0801, Japan}

\author{Akio Tomiya}
\email[]{akio@het.phys.sci.osaka-u.ac.jp}
\affiliation{Department of Physics, Graduate School of Science, Osaka University,
  Toyonaka, Osaka 560-0043, Japan}

\begin{abstract}
  We investigate the effects of the violation of the
  Ginsparg-Wilson (GW) relation in the M\"obius domain-wall fermion 
  formulation on the lattice with finite fifth dimension.
  Using a decomposion in terms of the eigenmodes of its
  four-dimensional effective Dirac operator, we isolate the 
  GW-violating terms for various physical quantities including the
  residual mass and the meson susceptibilities relevant for the
  effective restoration of the axial U(1) symmetry at finite temperature.
  Numerical result shows that the GW-violating effect is more
  significant, or even overwhelming, for the quantities that are
  dominated by the low-lying eigenmodes. 
\end{abstract}

\pacs{11.15.Ha, 
  11.30.Rd	
}

\maketitle

\section{Introduction}
Chiral symmetry is a notable property of the QCD lagrangian. Its
spontaneous breaking induces significant phenomenological
properties of low-energy QCD as described by chiral effective
theories.
In the lattice study of low-energy QCD it is therefore highly
desirable to maintain the chiral symmetry on the lattice to have
cleaner control of the chiral properties.
Indeed, there is a class of physical quantities that are sensitive even to
tiny violations of chiral symmetry.
A well-known example is the difference between vector and axial-vector
vacuum polarization functions $(\Pi_{VV}-\Pi_{AA})(Q^2)$, which 
vanishes in the limit of unbroken chiral symmetry.
In its numerical study on the lattice \cite{Shintani:2008qe}
chiral symmetry plays a crucial role.
Another important example is the difference of the meson
susceptibilities $\chi_\pi-\chi_\delta$, which is an order parameter
of the flavor-singlet axial $U(1)$ symmetry and it is used in the
study of effective restoration of axial $U(1)$ at finite
temperature (for recent lattice studies see \cite{Cossu:2013uua,Buchoff:2013nra,Chiu:2013wwa,Brandt:2013mba}).
Written in terms of fermion eigenmodes, it is a weighted average of
the Dirac eigenvalue spectrum $\rho(\lambda)$ with the dominant
contribution from the lowest end of the spectrum, and thus any small
effect of chiral symmetry violation on the near-zero Dirac eigenmodes
may substantially affect the result. 

In this work we study the effect of the violation of chiral symmetry
on the lattice with an application to the meson susceptibilities in
mind.
We consider a variant of the domain-wall fermion formulation \cite{Kaplan:1992bt,Shamir:1993zy,Furman:1994ky}, which
realizes chiral symmetry on the lattice very precisely but not
exactly, and formulate the remnant violation of chiral symmetry in the
basis of the Dirac eigenmodes.
Then, we are able to identify and monitor the violation for each
eigenmode and to trace its effect on the physical quantities such as
the meson susceptibilities.
In addition to the meson susceptibilities, we also analyze the
residual mass and the chiral condensate as simpler examples.

We use the lattice ensembles generated for a study of finite
temperature phase transition with the M\"obius domain-wall fermion. 
The physical results will be presented in a separate paper.

The paper is organized as follows. 
In Section~\ref{sec:Chiral} we first review the lattice fermion
formulation that we use in this study and then
decompose the meson susceptibilities and other quantities in the
Dirac eigenmodes basis in Section~\ref{sec:Meson}.
In Section~\ref{sec:Numerics} we show that
at finite temperature the contribution of the GW violation to the
susceptibilities on coarse lattices may dominate the signal. 
We repeat the same analysis for $m_\text{res}$ and show mode by mode
which are the major contributions of the GW violation term to the residual mass. 
We then summarize our results. 

\section{Chiral symmetry on the lattice}
\label{sec:Chiral}

\subsection{M\"obius domain-wall fermion}
Designing a discretized Dirac operator $D$ that preserves chiral
symmetry  without introducing unwanted doublers had been a long
standing problem in lattice field theory before the domain-wall \cite{Kaplan:1992bt,Shamir:1993zy,Furman:1994ky} and
overlap fermions \cite{Neuberger:1997fp,Neuberger:1998my} were discovered. 
Doublers inevitably appear on the lattice if we insist on the
continuum anti-commutation relation $\{\gamma_5,D\}=0$ for the lattice
Dirac operator with the generator of the chirality transformation
$\gamma_5$. 
This is known as the no-go theorem by Nielsen-Ninomiya
\cite{Nielsen:1981hk}. 
The solution to this conundrum was to modify the anti-commutator as
\begin{equation}
  \{\gamma_5, D\} = a R D\gamma_5 D,
\label{eq:GW}
\end{equation}
which is called the Ginsparg-Wilson (GW) relation
\cite{Ginsparg:1981bj}.
$R$ could be any local operator that commutes with $D^\dagger D$.
Any operator satisfying this relation has an exact chiral symmetry at
finite lattice spacing $a$~\cite{Luscher:1998pqa}. 
For a constant $R$, which we take $R=2$, a formal solution is known, 
{\it i.e.} the overlap Dirac operator 
\cite{Neuberger:1997fp,Neuberger:1998my}:
\begin{equation}
  D_{\rm ov} = \frac{1}{2} \left[
    1 + \gamma_5 \Sgn[H_K(M_5)
  \right],
\end{equation}
where the kernel $H_K$ is an hermitian Wilson-like operator 
$\gamma_5(D_K - M_5)$ with a large negative mass term $M_5$. 
Here and in the following we omit the lattice spacing for simplicity;
physical quantities are to be extracted after multiplying an
appropriate factor of $a$.
$D_{\rm ov}$ has all desirable properties of a chirally symmetric
operator at the cost of evaluating the matrix sign function, which 
is possible but the lattice size that could be treated with a given
machine time is much more limited than with other (non-chiral) lattice
fermion formulations.
For attempts of its large-scale simulation see for instance 
\cite{Aoki:2008tq, Borsanyi:2012xf}. 
A closely related formulation is the so-called domain-wall fermion
\cite{Kaplan:1992bt,Shamir:1993zy,Furman:1994ky}. 
By introducing a five-dimensional (5D) fermionic field with special 
boundary conditions, it realises a chiral fermion mode on 
its four-dimensional (4D) boundaries. 
An equivalence of the two formulations can be proved explicitly in the
limit of large fifth dimension $L_s \rightarrow \infty$ 
\cite{Narayanan:1992wx,Narayanan:1993ss,Narayanan:1993sk,Kikukawa:1999sy}. 

There are related lattice fermion formulations of the domain-wall class
other than its original form \cite{Furman:1994ky}.
They have different structures of 5D Dirac operators, generally
written as $D^{5d}_\text{DW}(m)$;
after constructing the 4D effective Dirac operator 
they are distinguished by the form of the kernel operator $H_K$
and the sign function approximation.
The 4D effective Dirac operator is constructed as
\begin{equation}
  D^{4d}_\text{DW}(m) = 
  \left[ \mathcal P^{-1}
    (D^{5d}_\text{DW}(1))^{-1} D^{5d}_\text{DW}(m)
    \mathcal P
  \right]_{11} ,
  \label{eq:D4d}
\end{equation}
where $\mathcal P$ is a permuation matrix defined in the 5D space 
to place the left-handed and right-handed modes of the physical 4D
field on the opposite 4D surfaces of the 5D space.
$D^{5d}_\text{DW}(1)$ is the Pauli-Villars regulator to kill
unnecessary degrees of freedom in the 5D space.
For the simple cases that the structure of the 5D operator does not
depends on the coordinate $s$ of the fifth direction,
the form of the 4D effective operator is given as
\begin{equation}
  D^{4d}_\text{DW}(m) =
  \frac{1+m}{2}+\frac{1-m}{2}\gamma_5\frac{T(H_K)^{-L_s}-1}{T(H_K)^{-L_s}+1} ,
  \label{eq:D4deff}
\end{equation}
with $T(H_K)=(1-H_K)(1+H_K)^{-1}$.

In the so-called M\"obius domain-wall fermion \cite{Brower:2012vk} 
the kernel is given by
\begin{equation}
  H_K(M_5) = \gamma_5 \frac{(b+c) D_W(M_5)}{2+ (b-c) D_W(M_5)},
\end{equation}
where $D_W$ is the standard Wilson-Dirac operator defined on
four-dimensional slices of the five-dimensional space. 
Two parameters $b$ and $c$ can be adjusted to control the chirality
properties. 
It includes two common choices as special cases:
(i) $b+c=2$, $b-c=0$, the Borici kernel~\cite{Borici:1999zw}, and 
(ii) $b+c=b-c=1$, the Shamir kernel~\cite{Shamir:1993zy}. 
One can also use a different setup $b+c=\alpha$, $b-c=1$,
which is called the scaled-Shamir kernel \cite{Brower:2012vk}. 
The scale factor $\alpha$ is to be chosen such that the sign
function approximation is optimized \cite{Brower:2012vk}.

At any finite $L_s$, the 4D effective operator $D^{4d}_\text{DW}(0)$
slightly violates the GW relation.
By writing 
\begin{equation}
  \{\gamma_5, D^{4d}_\text{DW}(0)\} 
  - 2 D^{4d}_\text{DW}(0)\gamma_5 D^{4d}_\text{DW}(0) 
  = 
  \gamma_5 \Delta ,
  \label{eq:GW2}
\end{equation}
$\Delta$ expresses the defect in the approximation of the sign
function 
\begin{equation}
  \Delta = \frac{1}{4}\left(1- \Sgn^2(H_K)\right).
  \label{eq:Defect}
\end{equation}
The effect of this small violation of chiral symmetry can be written
in terms of matrix elements of relevant operators including
$\Delta$. 
In the next section we consider such matrix elements in the space
spanned by the eigenmodes of hermitian operator
$H_m \equiv \gamma_5 D^{4d}_\text{DW}(m)$.
Their contributions to the physical quantities, such as the meson
susceptibilities and chiral condensate, are then studied
quantitatively on finite temperature lattices.

\subsection{Residual mass}
The residual mass $m_{\rm res}$ is a standard estimate of the GW violation
in the domain-wall fermion formulation. 
It represents the additive renormalization to the quark mass from the 
remnant violation of chiral symmetry at finite $L_s$.
Since one can write the axial-vector current $A_\mu$ and the
axial-Ward-Takahashi identity that it follows on the lattice 
\cite{Furman:1994ky,Boyle:2015vda},
$\Delta_\mu^- A_\mu^a = 2m P^a + 2J_{5q}^a$
with $\Delta_\mu^-$ the lattice (backward) derivative,
the residual mass is naturally defined through matrix elements such as
\begin{equation}
  m_{\rm res} = \frac{
    \langle 0| J_{5q}^a | \pi(\vec{p}=0)\rangle
  }{
    \langle 0| P^a | \pi(\vec{p}=0)\rangle
  },
  \label{eq:resmass_pi}
\end{equation}
where $P^a$ is the flavor non-singlet pseudo-scalar density and we
chose a zero-momentum pion as the external state.
The operator $J_{5q}^a$ is defined on the (unphysical) $L_s/2$-th slice
of the 5D space.
The calculation of $m_{\rm res}$ of this definition requires that the
operators $J_{5q}^a$ and $P^a$ are sufficiently separated from the source
to create the pion state to ensure the dominance of the ground state
pion. 

An alternative definition that does not refer to any external states
may be considered \cite{Brower:2012vk}: 
\begin{equation}
  m_{\rm res} = \frac{
    \left\langle {\rm Tr}\left[
        (\tilde D_m^{-1})^\dagger \Delta \tilde D_m^{-1}
      \right]
    \right\rangle
  }{
    \left\langle {\rm Tr} \left[
        (\tilde D_m^{-1})^\dagger \tilde D_m^{-1}
      \right]
    \right\rangle
  }.
  \label{eq:resmass_full}
\end{equation}
The massive quark propagator $\tilde{D}_m^{-1}$ is given by an
inverse of the 4D effective operator up to a contact term
\begin{equation}
  \tilde D_m^{-1}= \frac{1}{1-m}((D_{\rm DW}^{5d}(m))^{-1} - 1),
  \label{eq:PropSubtr}
\end{equation}
which coincides with the surface-to-surface domain-wall propagator. 
The denominator of (\ref{eq:resmass_full}) represents a pseudo-scalar
correlator from the origin to arbitrary space-time points summed over,
and thus contains the contribution from pion as well as those from
excited states.
In the numerator the defect operator $\Delta$ is inserted to probe the
GW violation.
It corresponds to the mid-point operator $J_{5q}^a$ rewritten by the 4D
effective field \cite{Brower:2012vk}.

\subsection{Meson susceptibilities and chiral condensate}
Meson susceptibilities are useful tools to probe the effective
restoration of the axial $U(1)$ symmetry at finite temperature.
In particular, we consider the connected pseudoscalar and scalar
susceptibilities $\chi_\pi-\chi_\delta$ that vanishes if the axial
symmetry $U(1)_A$ is restored. 
It is defined by
\begin{equation}
  \chi_\pi - \chi_\delta = 
  \frac{1}{V} \left\langle
    \int d^4x\ \pi^a(x) \pi^a(0) - 
    \int d^4x\ \delta^a(x) \delta^a(0)
  \right\rangle ,
  \label{eq:susc1}
\end{equation}
where $\pi^a(x) = i\bar\psi(x)\gamma_5\tau^a\psi(x)$,
$\delta^a(x) = \bar\psi(x)\tau^a\psi(x)$ 
and $V$ is the space-time volume. 
In terms of the massive quark propagator $\tilde{D}_m^{-1}$, 
they are written as
\begin{equation}
  \chi_\pi = \frac{1}{V} \langle\Tr [(\gamma_5 \tilde D_m)^{-2}]\rangle, 
  \quad 
  \chi_\delta = - \frac{1}{V} \langle\Tr[ (\tilde D_m)^{-2}]\rangle,
  \label{eq:SuscTraces}
\end{equation}
after averaging over the source point.
In the continuum theory the difference in (\ref{eq:susc1}) is written
in terms of the Dirac operator eigenvalue spectrum 
$\rho(\lambda)=(1/V)\langle\sum_{\lambda'}\delta(\lambda-\lambda')\rangle$
as
\begin{equation}
  \label{eq:suscCont}
  \chi_\pi-\chi_\delta = 
  \int_0^\infty d\lambda\,\rho(\lambda)
  \frac{4m^2}{(m^2+\lambda^2)^2},
\end{equation}
as a function of quark mass $m$.

Following the similar argument, the chiral condensate
$\langle\bar{\psi}\psi\rangle$ can also be constructed from the propagator,
\begin{equation}
  \langle\bar{\psi}\psi\rangle = 
  \frac{1}{V}\langle \Tr[(\tilde{D}_m)^{-1}] \rangle,
\end{equation}
and then in the continuum limit it is written by the spectral function
as 
\begin{equation}
  \langle\bar{\psi}\psi\rangle = 
  \int_0^\infty d\lambda\,\rho(\lambda)
  \frac{2m}{m^2+\lambda^2},
\end{equation}
which leads to the Banks-Casher relation
$\langle\bar{\psi}\psi\rangle=\pi\rho(0)$ 
in the thermodynamical limit \cite{Banks:1979yr}.
Because of the factor $4m^2/(m^2+\lambda^2)^2$ in 
(\ref{eq:suscCont}), the difference
$\chi_\pi-\chi_\delta$ is more sensitive to the lowest end of the
spectrum $\rho(\lambda\approx 0)$.
Having a good control of chiral symmetry is therefore of paramount
importance for the difference of the meson susceptibilities.

\section{Eigenmode decomposition with the Ginsparg-Wilson-type fermions}\label{sec:Meson}

Let us now discuss the decomposition of these observables in terms of
the eigenmodes of the M\"obius domain-wall Dirac operator at finite
$L_s$.
The operator $\Delta$ to represent the GW violation is explicitly
taken into account. 
The GW violating terms in the decomposition
only depend on the matrix elements of $\Delta$, 
but for the sake of simplicity we define several related quantities
during the derivation. 
In particular, in view of the numerical computation it is easier to 
define the violation in terms of the matrix elements of $\gamma_5$. 



\subsection{Domain-wall Dirac operator eigenmodes}
As a basis for the study of the GW violation, we take the 
eigenmodes of the hermitian Dirac operator 
$H_0 \equiv \gamma_5 D^{4d}_\text{DW}(0)$, or those of the 
massive operator 
$H_m= (1-m) H_0 + \gamma_5 m$.
The eigenmodes of the hermitian operator are numerically easier to
calculate compared to its non-hermitian conterpart.
The GW relation (\ref{eq:GW2}) is rewritten as
\begin{equation}
  \{\hat \gamma_5, H_0 \} = \Delta ,
  \label{eq:GWH}
\end{equation}
with $\hat\gamma_5\equiv\gamma_5-H_0$.

The $n$-th eigenvalue $\lambda_n$ and its corresponding eigenmode
$|\psi_n\rangle$ of $H_m$ satisfy 
\begin{equation}
  H_m |\psi_n\rangle = \lambda_n |\psi_n\rangle.
\end{equation}
A matrix element of $\gamma_5$ for these eigenmodes is then written as
\begin{equation}
  \langle\psi_n|\gamma_5|\psi_n\rangle 
  = 
  \frac{\lambda_n^2 + m}{\lambda_n(1+m)} + g_{nn},
  \label{eq:g5matrix}
\end{equation}
where we define the contribution from the GW violating operator
$\Delta$,
\begin{equation}
  g_{nn} \equiv 
  \frac{(1-m)^2}{2(1+m)\lambda_n} \langle\Delta\rangle_{nn}.
  \label{eq:gnn}
\end{equation}
Here and in the following we use a shorthand notation
$\langle\mathcal{O}\rangle_{mn} \equiv 
\langle\psi_m|\mathcal{O}|\psi_n\rangle$.

For the chiral condensate
$\Sigma = (1/V) \langle \Tr [\tilde{D}^{-1}]\rangle$,
the decomposition is straightforward using
$\tilde H_m^{-1} = \frac{1}{1-m}(H_m^{-1} - \gamma_5)$,
{\it i.e.}
\begin{equation}
  \Tr[\gamma_5\Htilde]
  = 
  \frac{1}{1-m} \sum_n \left[
    \frac{\langle\gamma_5\rangle_{nn}}{\lambda_n} - 1
  \right] 
  = 
  \frac{m}{1-m^2} \sum_n \frac{1-\lambda^2_n}{\lambda^2_n} +
  \frac{1}{1-m} \sum_n \frac{g_{nn}}{\lambda_n}.
  \label{eq:ChiralCond}
\end{equation}
The first term is the physical contribution that survives in the limit
of $L_s\to\infty$, while the second term describes the effect of
the GW violation.

\subsection{Decomposition of meson susceptibilities}
For the pion susceptibility
$\chi_{\pi} = (1/V)\langle \Tr [(\gamma_5 \tilde D_m)^{-2}] \rangle 
= (1/V) \langle \Tr [\Htildesq ]\rangle$,
the decomposition is written as
\begin{equation}
\begin{split}
  \Tr[\Htildesq]
  &= 
    \frac{1}{(1-m)^2} \sum_n
    \langle (H_m^{-1} - \gamma_5)(H_m^{-1} - \gamma_5) \rangle_{nn}
  \\
  & = 
    \frac{1}{1-m^2}\sum_n \frac{1-\lambda^2_n}{\lambda^2_n} -
    \frac{2}{(1-m)^2}\sum_n \frac{g_{nn}}{\lambda_n}.
  \label{eq:pi}
\end{split}
\end{equation}
The first term, that represent the result without any GW relation violation, satisfies 
$\Tr[\Htildesq]=(1/m)\Tr[\gamma_5\Htilde]$,
which corresponds to the relation 
$\chi_\pi = (1/m)\langle\bar{\psi}\psi\rangle$
derived from the axial Ward-Takahashi identity.

The calculation for the $\delta$ susceptibility
$\chi_{\delta} = -(1/V) \langle\Tr[\Dtildesq]\rangle 
= -(1/V)\langle\Tr[(\gamma_5\tilde H_m)^{-2}]\rangle$
requires another set of matrix elements. 
The first step is
\begin{equation}
\begin{split}
  \Tr[(\gamma_5 \tilde H_m)^{-2}]
  &= 
  \frac{1}{(1-m)^2} \sum_n
  \langle (\gamma_5 H_m^{-1} - 1)(\gamma_5 H_m^{-1} - 1) \rangle_{nn}
  \\
  &= \frac{1}{(1-m)^2} \sum_n \left[
    \frac{1}{\lambda_n} \langle\gamma_5H_m^{-1}\gamma_5\rangle_{nn} -
    \frac{2}{\lambda_n} \langle\gamma_5\rangle_{nn} + 1 
  \right],
  \label{eq:delta1}
\end{split}
\end{equation}
where an expression for 
$\langle\gamma_5H_m^{-1}\gamma_5\rangle_{nn}$ is needed. 
This can be decomposed into the terms satisfying and violating the GW
relation: 
\begin{equation}
  \langle\gamma_5H_m^{-1}\gamma_5\rangle_{nn} = 
  \frac{1}{\lambda_n} \left[ 
    2 \left(\frac{\lambda_n^2+m}{\lambda_n(1+m)}\right)^2 -1 
  \right] + h_{nn}.
  \label{eq:G5HG5}
\end{equation}
which defines a new quantity, $h_{nn}$, as the defect from the violation of
the GW relation. 
It contains the diagonal elements of $\langle\Delta\rangle_{nk}$ as
well as its off-diagonal components through
\begin{equation}
  \begin{split}
    \langle\gamma_5H_m^{-1}\gamma_5\rangle_{nn} &= 
    \sum_{k} \frac{1}{\lambda_k} |\langle\gamma_5\rangle_{nk}|^2 ,
    \\
    (1-m)^2\langle \Delta \rangle_{nk} &= \langle \gamma_5 \rangle_{nk} (1+m) (\lambda_n + \lambda_k) - 2\delta_{nk} (\lambda_n^2 + m).
\label{eq:mixedgamma5}
\end{split}
\end{equation}
Using these, we obtain
\begin{equation}
\begin{split}
  \Tr[(\gamma_5\tilde H_m)^{-2}] 
  &= 
    \frac{1}{(1-m^2)^2} \sum_n \frac{1-\lambda_n^2}{\lambda_n^4}
    \left[2m^2 - \lambda_n^2(1+m^2) \right]
  \\
  &+ \frac{1}{(1-m)^2} \sum_n 
    \left[\frac{h_{nn}}{\lambda_n} - \frac{2g_{nn}}{\lambda_n}\right],
\end{split}
\end{equation}
where the violation of the GW relation is isolated in the second term.

The difference 
$\langle\Delta_{\pi-\delta}\rangle \equiv
\chi_\pi-\chi_\delta$ can then be written using
\begin{equation}
\begin{split}
  \Delta_{\pi-\delta} &=
  \frac{1}{V} (\Tr[\Htildesq] + \Tr[(\gamma_5\tilde H_m)^{-2}])
  \\
  &= \frac{1}{V(1-m^2)^2} \sum_n
  \frac{2m^2(1-\lambda_n^2)^2}{\lambda_n^4}
  +\frac{1}{V(1-m)^2} \sum_n 
  \left[\frac{h_{nn}}{\lambda_n}-\frac{4g_{nn}}{\lambda_n}\right].
  \label{eq:DeltaViol}
\end{split}
\end{equation}
For later use, we define 
$\Delta_{\pi-\delta}^{\rm GW}$ and
$\Delta_{\pi-\delta}^{\cancel{\rm GW}}$ for the first and second term of
the right hand side of this equation.
Contrary to the individual susceptibilities $\chi_\pi$ and
$\chi_\delta$, that have a quadratic ultraviolet divergence,
the difference has only a logarithmic divergence in 
$\Delta_{\pi-\delta}^{\rm GW}$.
Notice also that the maximum eigenvalue of the domain-wall effective Dirac
operator is one, and the numerator $(1-\lambda_n^2)^2$ is highly
suppressed near the ultraviolet end.
Indeed, we can confirm reasonable saturation by summing $O(10)$ 
low-lying eigenmodes as discussed in the next section.

\subsection{Decomposition of residual mass}
In order to obtain the eigenvalue decomposition of the residual mass (\ref{eq:resmass_full}),
we may use
\begin{equation}
\begin{split}
  \Tr[\tilde H_m^{-1} \Delta \tilde H_m^{-1}] 
  &= 
    \frac{1}{(1-m)^2} \sum_n\left[
    \frac{\langle\Delta\rangle_{nn}}{\lambda_n^2} 
    - \frac{2\langle\Delta\gamma_5\rangle_{nn}}{\lambda_n} 
    + \langle\gamma_5\Delta\gamma_5\rangle_{nn} 
    \right]
  \\
  &= \frac{1}{(1-m)^2} \left[
    \sum_n \frac{1+\lambda_n^2}{\lambda_n^2} \langle\Delta\rangle_{nn}
    - \sum_n \frac{2\langle\Delta\gamma_5\rangle_{nn}}{\lambda_n}
    \right].
\end{split}
\end{equation}
The last equation is exact when we sum over all eigenmodes. 
The matrix element in the first term $\langle\Delta\rangle_{nn}$ is
proportional to $g_{nn}$ as in \eqref{eq:gnn}.
The second part, coming from the contact term in the Dirac propagator,
includes the off-diagonal elements $\langle\gamma_5\rangle_{nk}$ and
thus depends also on the off diagonal elements $\Delta_{nk}$, like
$h_{nn}$ does.
By inserting the complete set of eigenmodes it may also be written as
\begin{equation}
  (1-m)^2\langle\Delta\gamma_5\rangle_{nn} 
  = (1+m) \sum_k |\langle\gamma_5\rangle_{kn}|^2 
  (\lambda_n +\lambda_k)
  - 2 \langle\gamma_5\rangle_{nn}(\lambda_n^2+m),
\end{equation}
which is used to evaluate $\langle\Delta\gamma_5\rangle_{nn}$ with the
low-lying eigenmodes.

The eigenmode decomposition of the residual mass reads
\begin{equation}
  m_{\rm res} = 
  \frac{
    \left\langle
      \displaystyle
      \sum_n \left[
        \frac{1+\lambda_n^2}{\lambda_n^2} \langle\Delta\rangle_{nn} -
        \frac{2}{\lambda_n} \langle\Delta\gamma_5\rangle_{nn}
      \right]
    \right\rangle
  }{
    \left\langle
      \displaystyle
      \sum_n \left[
        \frac{1+\lambda_n^2}{\lambda_n^2} - 
        \frac{2}{\lambda_n} \langle\gamma_5\rangle_{nn}
      \right]
    \right\rangle
  }.
  \label{eq:ResMassEmodes}
\end{equation}
The sum on the denominator leads to an ultraviolet divergence, given
the asymptotic scaling of the spectral function
$\rho(\lambda)\sim\lambda^3$.
Also, the numerator could diverge unless $\langle\Delta\rangle_{nn}$
and $\langle\Delta\gamma_5\rangle_{nn}$ dump rapidly at high energies.
We can still {\it define} the residual mass with some fixed number of
the eigenmodes or at a fixed cutoff on $\lambda$, that works as a
regularization for this particular quantity.
The residual mass thus defined may probe the GW violation at low
energies, while the one with all modes is dominated by the
contributions of eigenmodes of order of the lattice cutoff.

\section{Numerical results}\label{sec:Numerics}
In this section we show the numerical results obtained on finite
temperature lattices, for which the chiral condensate and meson
susceptibilities play a crucial role to characterize the property of
the phase transition.

\subsection{Lattice setup}
We perform numerical simulations of two-flavor QCD at finite
temperatures near the critical temperature $T_c\approx$ 170--180~MeV.
The temporal extent $N_t$ is either 8 or 12, and the spatial dimension
is $N_s=16$ or 32 for $N_t=8$ and $N_s=32$ for $N_t=12$.
We use the tree-level Symanzik improved gauge action and the M\"obius
domain-wall fermion action.
For the M\"obius domain-wall fermion, we set $L_s$ = 12, 16 or 24,
depending on ensembles.
In the following we denote the lattice size in the format 
$N_s^3\times N_t\,(\times L_s)$.
The three-step stout link-smearing \cite{Morningstar:2003gk} is
introduced for the link variables in the domain-wall fermion action,
that helps to reduce the residual mass with the modest values of $L_s$ 
mentioned above.
A wide range of (degenerate) quark masses is taken for some of the
ensembles. 
The residual mass we observe with various definitions is discussed
later in the section.

\begin{table}[tbp]
  \begin{center}
    \begin{tabular}{cl c c rr cc}
      \hline
      $\beta$ & $m$ & $N_s^3\times N_t\,(\times L_s)$ & $a$ (fm)
      & $N_{\rm conf}$ & $N_{\rm ev}$ 
      & $\langle\Delta_{\pi-\delta}^{\cancel{\rm GW}}\rangle /
         \langle\Delta_{\pi-\delta}^{(N_{\rm ev})}\rangle$
      & $\langle\Delta_{\pi-\delta}\rangle /
         \langle\Delta_{\pi-\delta}^{(N_{\rm ev})}\rangle$ \\\hline 
      4.07	&   0.01	&$16^3\times 8\,(\times 12)$ & 0.121  &	239& 86&    0.378 $\pm$	0.026	&	1.0015 $\pm$          0.0090    \\
      4.10	&   0.01        &$16^3\times 8\,(\times 12)$ & 0.113  &	203& 86&    0.279 $\pm$ 0.040   &	0.9874 $\pm$          0.0117    \\
      4.10	&   0.01        &$32^3\times 8\,(\times 12)$ & 0.113  &         85& 48&    0.302 $\pm$ 0.024   &	1.0074 $\pm$          0.0057    \\
      4.07	&   0.001       &$16^3\times 8\,(\times 24)$ & 0.121  &	210& 69&    0.982 $\pm$ 0.002   &	1.0124 $\pm$          0.0198    \\
      4.07	&   0.001       &$32^3\times 8\,(\times 24)$ & 0.121  &	215& 25&    0.654 $\pm$ 0.105   &	0.9979 $\pm$          0.0209    \\
      4.10	&   0.001       &$16^3\times 8\,(\times 24)$ & 0.113  &	124& 84&    0.983 $\pm$	0.004	&	0.9937 $\pm$	      0.0162    \\
      4.10	&   0.005       &$32^3\times 8\,(\times 24)$ & 0.113  & 	 98& 28&    0.170 $\pm$	0.023	&	1.0090 $\pm$	      0.0113    \\
      4.10	&   0.001       &$32^3\times 8\,(\times 24)$ & 0.113  & 	 69& 36&    0.975 $\pm$	0.006	&	1.0220 $\pm$	      0.0108    \\
      4.18	&   0.01        &$32^3\times 8\,(\times 12)$ & 0.096  & 	 80& 48&    0.616 $\pm$	0.061	&	1.0969 $\pm$	      0.0279    \\\hline
      4.18	&   0.01	&$32^3\times 12\,(\times 16)$& 0.096  &	 54& 15&    0.080 $\pm$	0.006	&	1.0126 $\pm$	      0.0043    \\
      4.22	&   0.01	&$32^3\times 12\,(\times 16)$& 0.088  &	 50& 38&    0.053 $\pm$	0.006	&	1.0051 $\pm$	      0.0069    \\
      4.23	&   0.01	&$32^3\times 12\,(\times 16)$& 0.086  &       94& 22&    0.038 $\pm$	0.004	&	1.0020 $\pm$	      0.0056    \\
      4.23	&   0.005	&$32^3\times 12\,(\times 16)$& 0.086  &      176& 30&    0.083 $\pm$	0.009	&       1.0054 $\pm$	      0.0064    \\
      4.23	&   0.0025	&$32^3\times 12\,(\times 16)$& 0.086  &       78& 24&    0.162 $\pm$	0.022	&       1.0028 $\pm$	      0.0119    \\
      4.24	&   0.01	&$32^3\times 12\,(\times 16)$& 0.084  &	326& 38&    0.046 $\pm$	0.003	&	1.0077 $\pm$	      0.0025    \\
      4.24	&   0.005	&$32^3\times 12\,(\times 16)$& 0.084  &       95& 33&    0.057 $\pm$	0.011	&       0.9883 $\pm$	      0.0077    \\
      4.24	&   0.0025	&$32^3\times 12\,(\times 16)$& 0.084  &      111& 39&    0.323 $\pm$	0.080   &       0.9995 $\pm$	      0.0214    \\
      4.30	&   0.01	&$32^3\times 12\,(\times 16)$& 0.075  &      195& 38&    0.007 $\pm$	0.003	&	1.0170 $\pm$	      0.0060    \\
      \hline
    \end{tabular}
    \caption{
      Lattice ensembles used in this study.
      The lattices of $N_t=8$ (=12) are listed in the first (second)
      block.
      In the last two columns, the fraction of the GW violating
      contribution to $\langle\Delta_{\pi-\delta}\rangle$ as well as
      of the partial sum up to $N_{\rm ev}$ eigenmodes
      $\langle\Delta_{\pi-\delta}^{(N_{\rm ev})}\rangle$ 
      are listed.   
    }
    \label{tab:ensembles}
  \end{center}
\end{table}

The details of the lattice ensembles used in this analysis are summarized in
Table~\ref{tab:ensembles}.
We use the standard Hybrid Monte Carlo (HMC) algorithm with multiple integration levels. 
The simulation code is described in~\cite{Cossu:2013ola}. 
Unlike the previous work with the overlap fermion formulation
\cite{Cossu:2013uua}, we do not introduce the term to prevent the
change of global topology \cite{Fukaya:2006vs}.
In fact we observe frequent topology tunneling events on these
ensembles. 
These lattices are used in the study of the effective restoration of
the axial $U(1)$ symmetry. Preliminary results were presented in 
\cite{Cossu:2014aua,Tomiya:2014mma}. 

On these ensembles we calculate $N_{\rm ev}$ lowest-lying
eigenvalues and eigenvectors of the hermitian 4D effective Dirac
operator $H_m$.
We use the implicitly restarted Lanczos algorithm to numerically
obtain the eigenmodes.
For the 4D effective operator, we need an inversion of the
Pauli-Villars regulator $(D_{\rm DW}^{5d}(1))^{-1}$ for each
application of $H_m$.
The calculation of the eigenmodes is done for configurations separated
by at least 20 trajectories of HMC (typically 50).

\subsection{GW violation for individual eigenmodes}

\begin{figure}[tbp]
  \includegraphics[width=.4\textwidth]{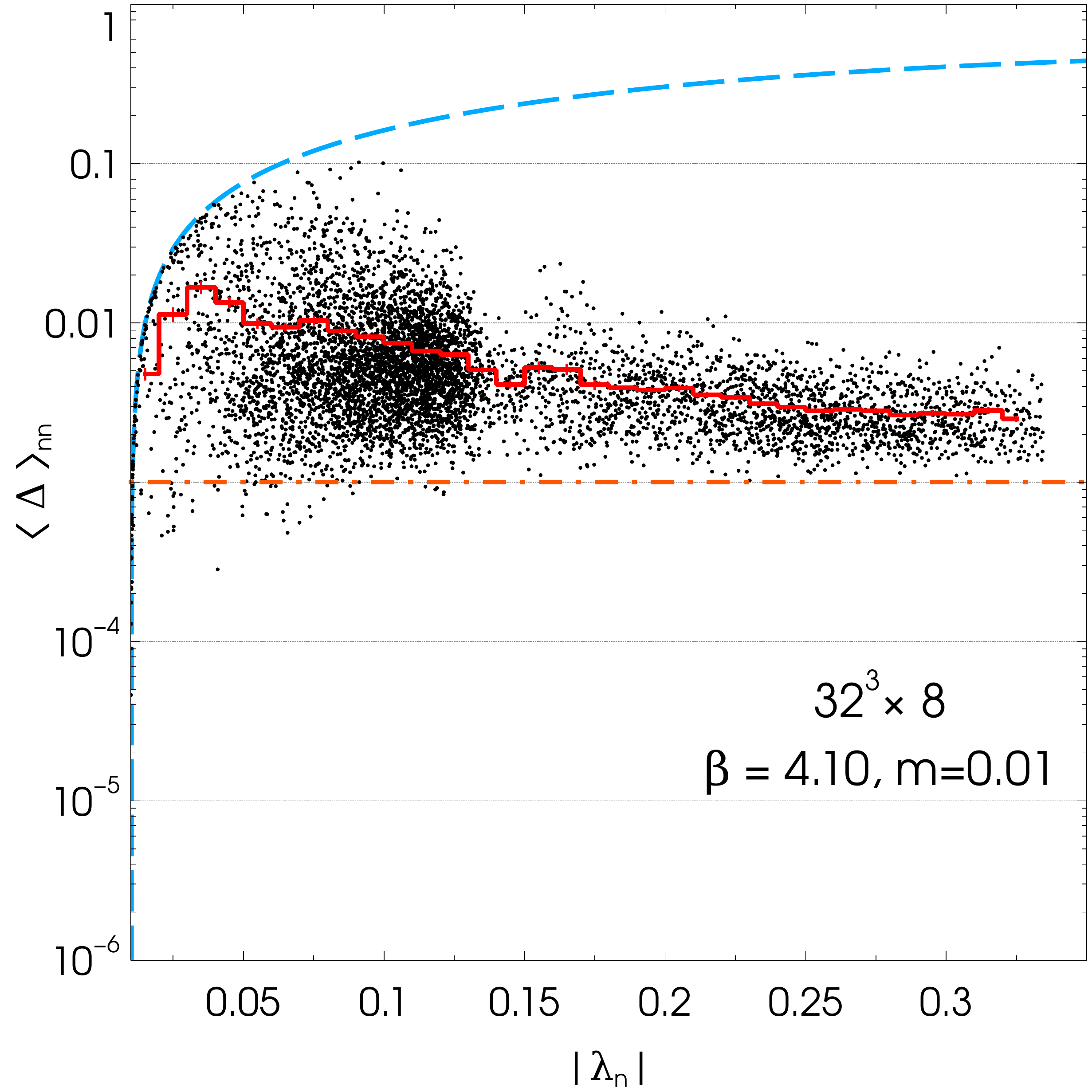}
  \hspace*{0.02\textwidth}
  \includegraphics[width=.4\textwidth]{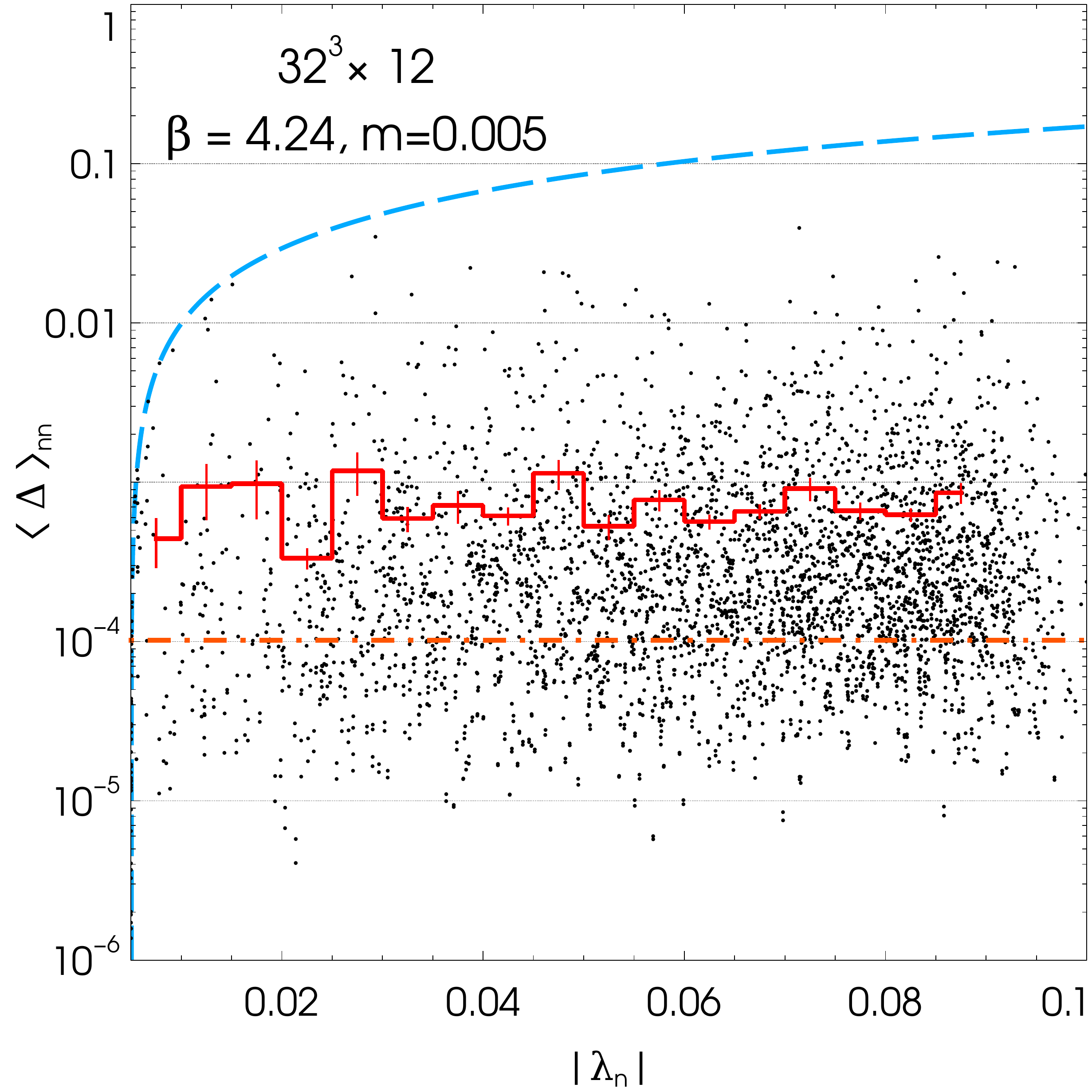}
  \caption{
    Scatter plot of $\langle\Delta\rangle_{nn}$ versus $\lambda_n$
    at $\beta=4.10$, $32^3\times 8$ and $m=0.01$ (left), and
    at $\beta=4.24$, $32^3\times 12$ and $m=0.005$ (right).
    The red thick line shows a binned average in a given range of
    $\lambda$.
    The blue dashed curve shows the maximal possible violation 
    $2(\lambda - m)(1-\lambda)$,
    and the orange dash-dot line is the result of the stochastic
    measurement of the residual mass.
  }
  \label{fig:DeltaBin}
\end{figure}

Figure~\ref{fig:DeltaBin} shows scatter plots of 
$\langle\Delta\rangle_{nn}$ versus $|\lambda_n|$
for the ensembles of 
$\beta=4.10$, $32^3\times 8\,(\times 12)$ at $m=0.01$ (left) and
$\beta=4.24$, $32^3\times 12\,(\times 16)$ at $m=0.005$ (right).
For individual eigenvalues, the matrix element
$\langle\Delta\rangle_{nn}$ takes the values between roughly 
$10^{-3}$ and 0.1 at $\beta=4.10$, and between
$10^{-5}$ and 0.01 at $\beta=4.24$ (right).
The maximum possible value of $\langle\Delta\rangle_{nn}$ occurs when 
$\langle\gamma_5\rangle_{nn}=1$ and is given by
$2(\lambda_n - m)(1-\lambda)$, which is shown in the plots by a blue
dashed curve.

From Figure~\ref{fig:DeltaBin} we clearly observe that the overall
size of $\langle\Delta\rangle_{nn}$ is made smaller for a finer
lattice.
The average value calculated in a bin of $|\lambda_n|$ shows a
reduction of an order of magnitude from $\beta=4.10$ to 4.24.
This is expected because the violation of chiral symmetry should
vanish in the continuum limit.

More importantly, the average value for a given range of $\lambda_n$
gradually decreases for larger $|\lambda_n|$.
Also shown by a thick dot-dashed line is the corresponding stochastic
estimate of the residual mass taking account of all the eigenmodes.
The binned average in a given range of $|\lambda_n|$ shown by red
lines indeed shows a tendency approaching the residual mass at large 
$\lambda_n$'s. (The available range of $\lambda$ is too narrow on the right
to see the decrease.)
Nearly maximum violation (blue curve) is observed only for near-zero
modes.

\begin{figure}
  \includegraphics[width=.95\textwidth]{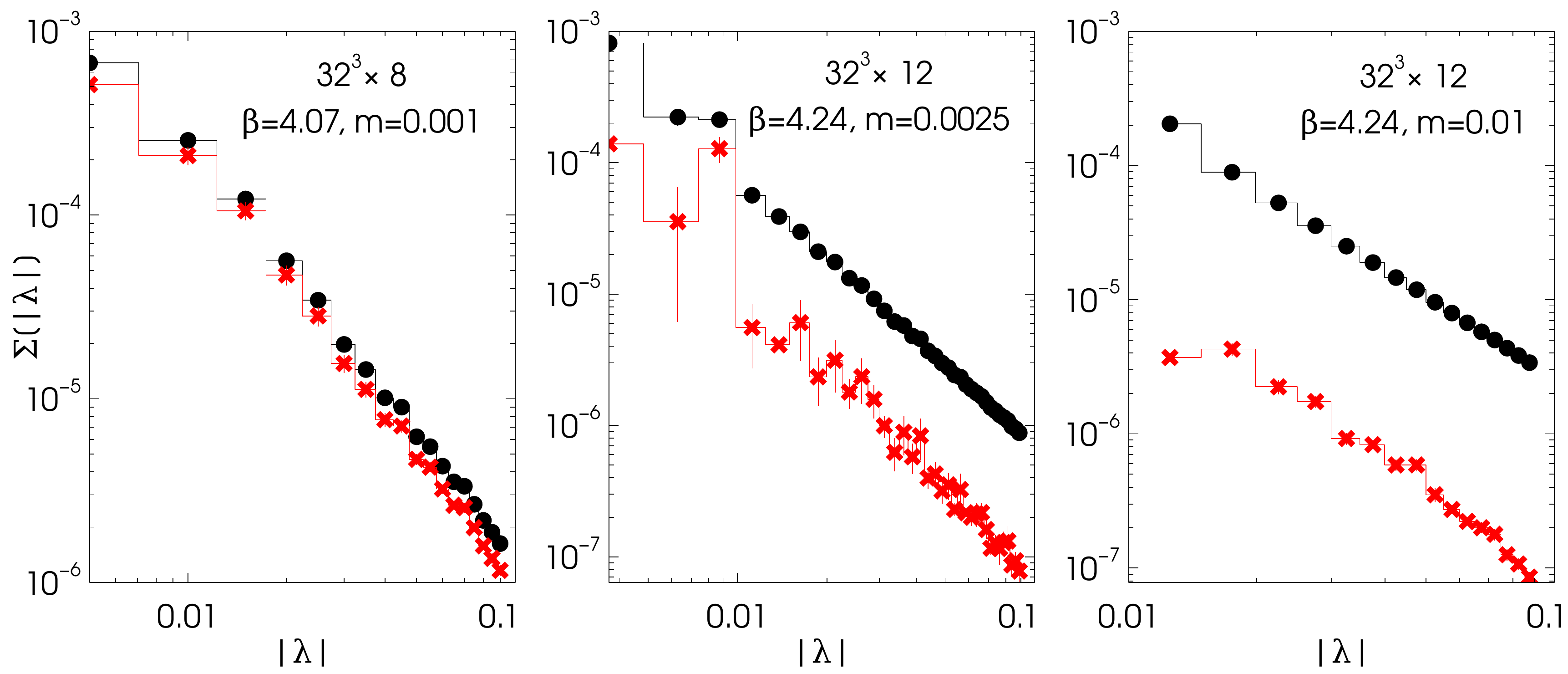}
  \caption{
    Contribution of the GW violating term to the chiral
    condensate averaged in small ranges of $|\lambda|$.
    Black histogram represents the full sum in (\ref{eq:ChiralCond})
    while red crosses show the GW violation term.
    The data at
    $\beta=4.07$, $32^3\times 8\,(\times 12)$, $m=0.001$ (left panel),
    $\beta=4.24$, $32^3\times 12\,(\times 16)$, $m=0.0025$ (middle),
    and
    $\beta=4.24$, $32^3\times 12\,(\times 16)$, $m=0.01$ (right panel)
    are shown.
  }
  \label{fig:ChiralCond}
\end{figure}

The $\langle\Delta\rangle_{nn}$ also appears in the decomposition
formula of chiral condensate (\ref{eq:ChiralCond}).
We study the contribution of the GW violating term
$\Sigma^{\cancel{\rm GW}}\equiv
(1/V(1-m))\langle\sum_n(g_{nn}/\lambda_n)\rangle$ 
to the full result by breaking the sum into small ranges of
$|\lambda|$.
In each bin of $|\lambda|$, we take an ensemble average and define
$\Sigma[|\lambda|]$ as well as $\Sigma^{\cancel{\rm GW}}[|\lambda|]$.
Figure~\ref{fig:ChiralCond} shows them as a function of $|\lambda|$ 
for comparable temperatures.

The plots demonstrates that the GW violating contribution
$\Sigma^{\cancel{\rm GW}}[|\lambda|]$
to $\Sigma[|\lambda|]$
is nearly dominating the entire sum on the $N_t=8$ lattice (left
panel). 
For most of the low-energy bins, it accounts for more than 50\% of the
total. 
For the finer lattice, $N_t=12$, the violating term 
$\Sigma^{\cancel{\rm GW}}[|\lambda|]$ becomes 5\% or less, when quark
mass is large $m=0.01$ (right panel).
The situation is worse at lower quark mass $m=0.0025$ (middle)
especially for the lowest-lying bins.

\subsection{Meson susceptibilities}
In this subsection we analyse the meson susceptibilities using the
decomposition formulae described in the previous section.

First of all, we emphasize that the difference of the meson
susceptibilities, $\langle\Delta_{\pi-\delta}\rangle$, is highly
dominated by the near-zero eigenmodes, because of a strong weight 
$1/\lambda_n^4$ for small eigenvalues as can be seen in the
decomposition formula (\ref{eq:DeltaViol}).
In particular, the exact zero-mode of $|\lambda_0|=m$ gives a
significant contribution on relatively small lattices, since the first
non-zero eigenvalue is well separated by $\sim 1/\Sigma V$
in the broken phase with the chiral condensate $\Sigma$.
In the unbroken phase, the non-zero modes are typically even more
separated.

\begin{figure}
  \includegraphics[width=.9\textwidth]{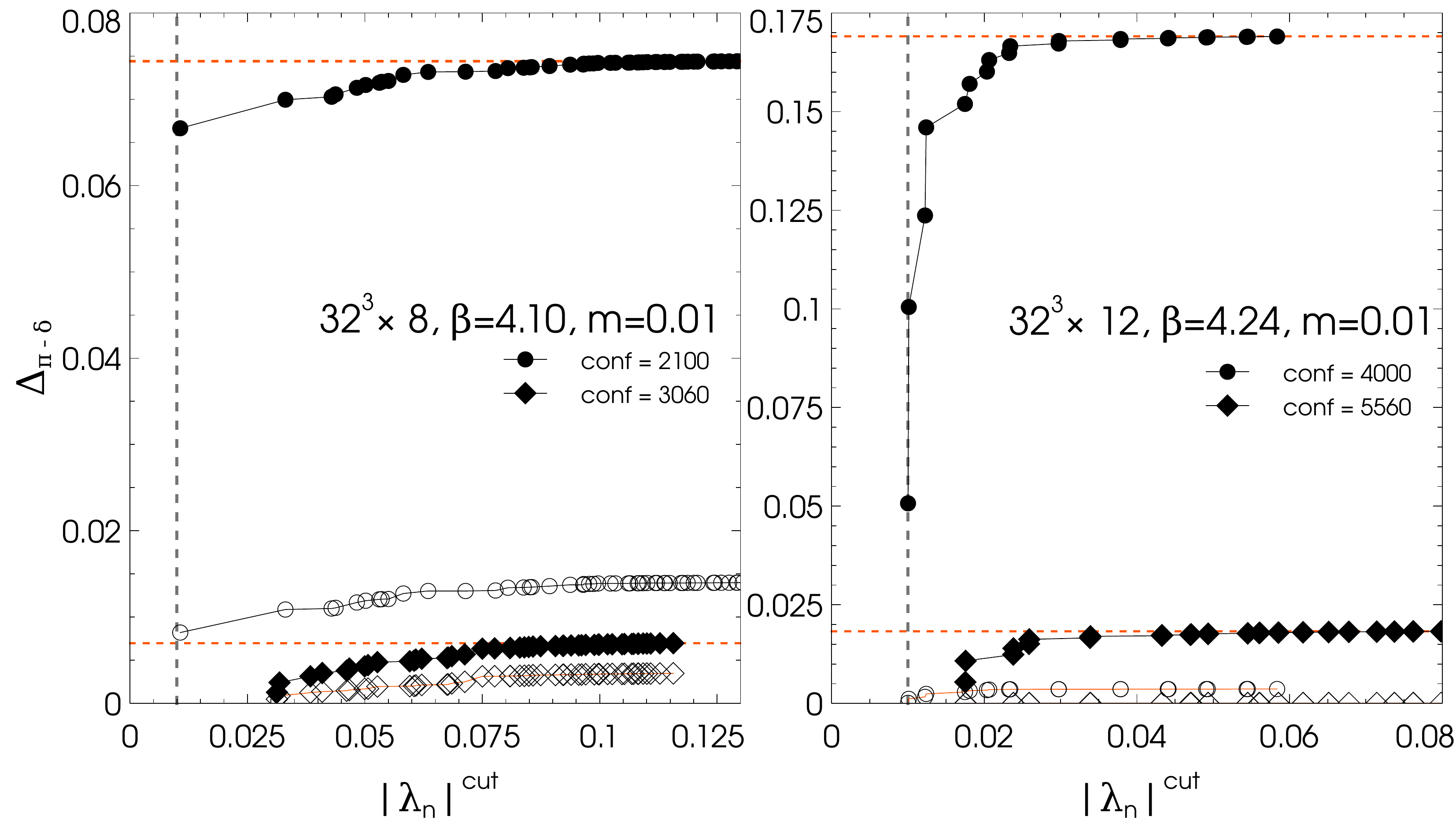}
  \caption{
    Accumulated contribution of the low-lying eigenmodes to 
    $\langle\Delta_{\pi-\delta}\rangle$.
    Examples for a couple of gauge configurations are shown for two 
    ensembles:
    $\beta=4.10$, $32^3\times 8\,(\times 12)$, $m=0.01$ (left) and 
    $\beta=4.24$, $32^3\times 12\,(\times 16)$, $m=0.01$ (right).
    Circles represent the gauge configurations with exact zero modes 
    $|\lambda_0|=m$, while diamonds correspond to those without
    zero-modes. Empty symbols are the corresponding contribution 
    from the violation terms.
    The dashed vertical line shows the position of a zero-mode
    $|\lambda_0|=m$.
    Horizontal lines are the full calculation without the cut, which
    is obtained by a stochastic method.
  \label{fig:Convergence}
  }
\end{figure}

A few examples are shown in Figure~\ref{fig:Convergence}, where
the contribution to $\Delta_{\pi-\delta}$ 
from low-lying eigenmodes below a cut $|\lambda_n|^{cut}$ is plotted
as a function of $|\lambda_n|^{cut}$.
We choose a couple of typical gauge configurations with and without exact
zero modes.
We find that the sum up to $|\lambda_n|^{cut}$ reaches a plateau at
relatively low values of
$|\lambda_n|^{cut}$, $\sim 0.1$ for the lattices of
$\beta=4.10$, $32^3\times 8\,(\times 12)$, $m=0.01$ (left), and 
$\sim 0.04$ for
$\beta=4.24$, $32^3\times 12\,(\times 16)$, $m=0.01$ (right).

This plot also shows a significant difference between the gauge
configurations with and without the exact zero-modes.
The configurations with a zero-mode (circles) give much larger value
compared to those without (diamonds).
The relative factor is 5--10, depending on the ensembles.

The effect of the GW violating term 
$\Delta_{\pi-\delta}^{\cancel{\rm GW}}$
is shown in the same plot by empty symbols.
The saturation by low-lying eigenmodes is seen also for this part of 
$\Delta_{\pi-\delta}$.

\begin{figure}
  \includegraphics[width=.8\textwidth]{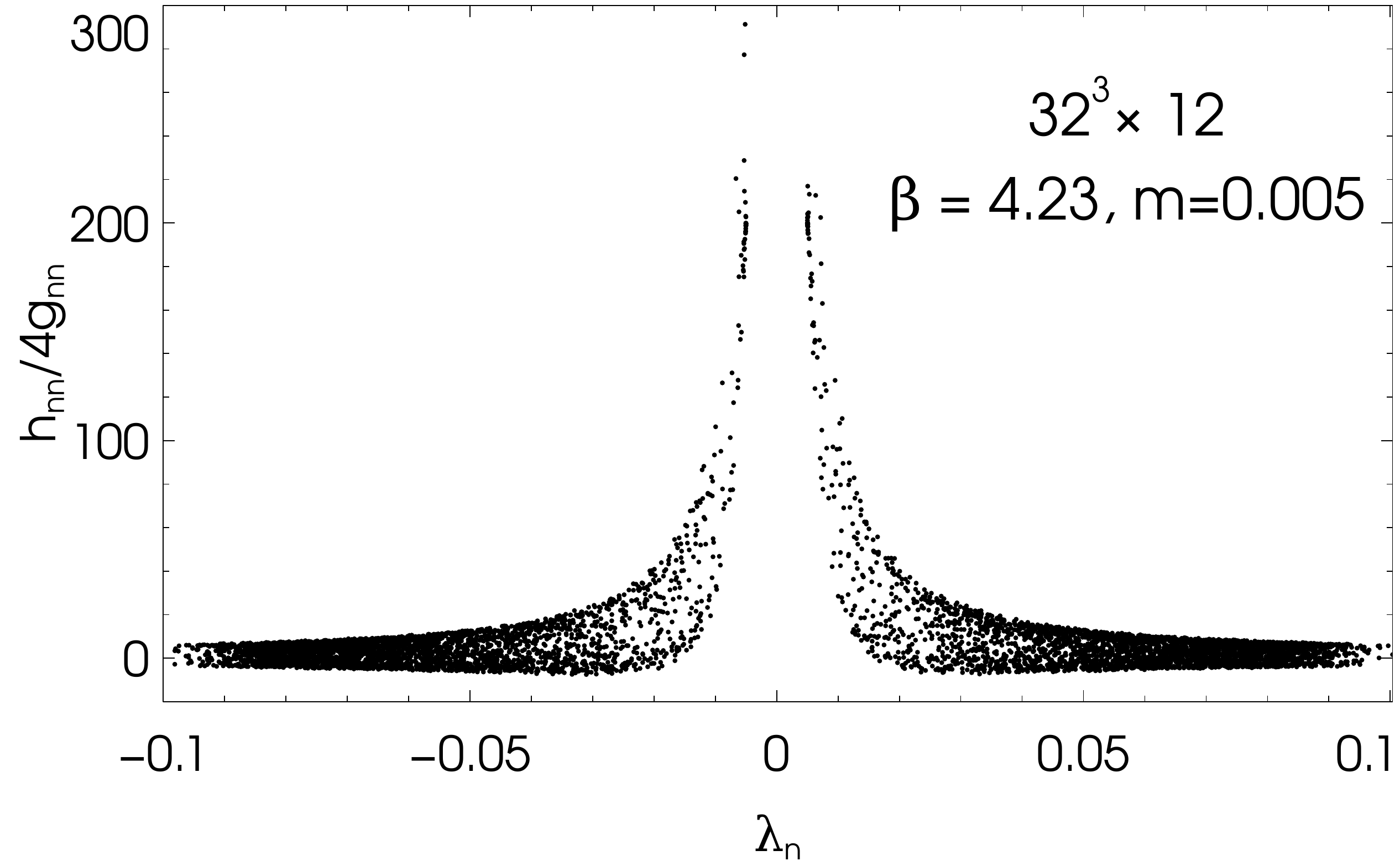}
  \caption{
    Ratio $h_{nn}/4g_{nn}$ for each eigenvalue $\lambda_n$.
    Data at $\beta=4.23$, $32^3\times 12(\times L_s=16)$, $m=0.005$
    are shown.
  \label{fig:HGratio}
  }
\end{figure}

In $\Delta_{\pi-\delta}^{\cancel{\rm GW}}$ there are two terms containing
$h_{nn}$ and $4g_{nn}$, respectively, as found in
(\ref{eq:DeltaViol}). 
The latter is proportional to $\langle\Delta\rangle_{nn}$ while the
former needs to be calculated from 
$\langle\gamma_5 H_m^{-1}\gamma_5\rangle_{nn}$, see \eqref{eq:G5HG5}.
It turned out that $h_{nn}$ dominates
$\Delta_{\pi-\delta}^{\cancel{\rm GW}}$ over $g_{nn}$ as
Figure~\ref{fig:HGratio} shows.
Here, the ratio of the $h_{nn}$ to $4g_{nn}$ is plotted for each
eigenmode.
The enhancement of $h_{nn}$ near $|\lambda_{n}|\simeq 0$ is very steep, and
dominates the sum 
$\Delta_{\pi-\delta}^{\cancel{\rm GW}}$
by a small number of low-lying modes.

On the two rightmost columns of Table~\ref{tab:ensembles} we summarize
the numerical results for
$\langle\Delta_{\pi-\delta}^{\cancel{\rm GW}}\rangle /
 \langle\Delta_{\pi-\delta}^{(N_{\rm ev})}\rangle$
and
$\langle\Delta_{\pi-\delta}\rangle /
 \langle\Delta_{\pi-\delta}^{(N_{\rm ev})}\rangle$.
The last column demonstrates the saturation of
$\langle\Delta_{\pi-\delta}\rangle$ with a limited number 
($N_{\rm ev}$) of low-lying eigenmodes,
which is denoted by $\langle\Delta_{\pi-\delta}^{(N_{\rm ev})}\rangle$,
for all ensembles we studied.
The full calculation is obtained using
the stochastic method with 15 $Z_2$ noise vectors. 
We confirm that this ratio is always consistent with unity.

The second last column, on the other hand, shows the fractional size
of the GW violating contribution 
$\langle\Delta_{\pi-\delta}^{\cancel{\rm GW}}\rangle$
to 
$\langle\Delta_{\pi-\delta}^{(N_{\rm ev})}\rangle$.
It shows larger variation between 0 and 1 depending on the ensemble. 
In the following, we discuss on some remarkable observations.

\begin{figure}[tbp]
  \includegraphics[width=.95\textwidth]{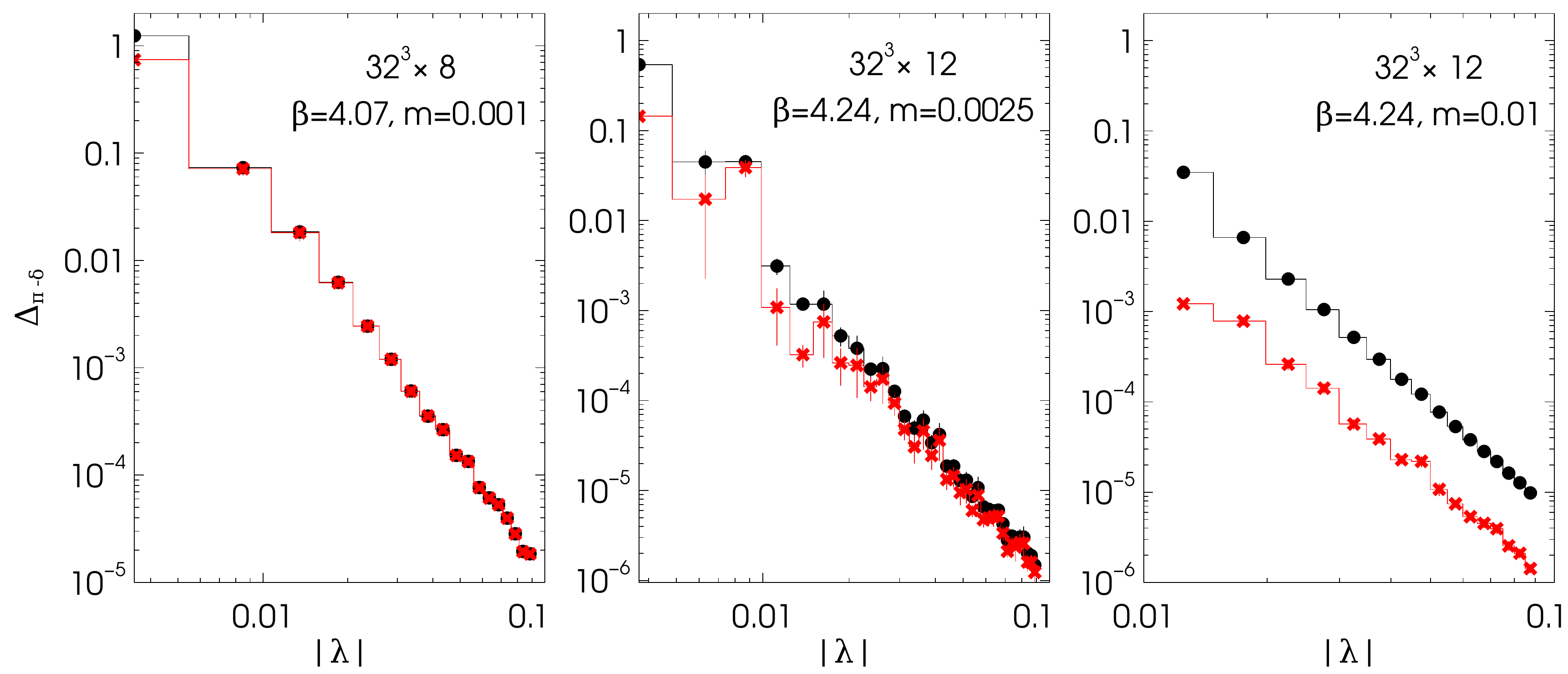}
  \caption{
    Eigenmode decomposition 
    of the suseptibility difference $\Delta_{\pi-\delta}$.
    Average in each bin of $\lambda$ is plotted for full contribution 
    (black circles) and for the contribution from the GW violating
    term $\Delta_{\pi-\delta}^{\cancel{\rm GW}}$ (red crosses).
  }
  \label{fig:DeltaHnn}
\end{figure}

Figure~\ref{fig:DeltaHnn} shows the size of the GW violating
contribution $\Delta_{\pi-\delta}^{\cancel{\rm GW}}$
in the eigenmode decomposition of 
$\Delta_{\pi-\delta}$.
For each bin of $|\lambda|$, we plot the average of full contribution
$\Delta_{\pi-\delta}$ (black circles) and a partial contibution from
the GW violating term $\Delta_{\pi-\delta}^{\cancel{\rm GW}}$.
We find that on the coarse lattice 
(left panel: $\beta=4.07$, $32^3\times 8\,(\times 24)$, $m=0.001$) 
the GW violating term gives a large
fraction, one third of total, at the lowest bin, and it even dominates
the signal for other bins.
The situation is better on the fine lattice when quark mass is large
(right panel: $\beta=4.24$, $32^3\times 12\,(\times 16)$, $m=0.01$).
Namely, the GW violating term is at least one order of magnitude
smaller than the total for all the measured bins.
Reducing the quark mass 
(middle panel: $\beta=4.24$, $32^3\times 12\,(\times 16)$,
$m=0.0025$), the GW violating contribution becomes more significant
especially for the lowest bins.

\begin{figure}[tbp]
  \includegraphics[width=.6\textwidth]{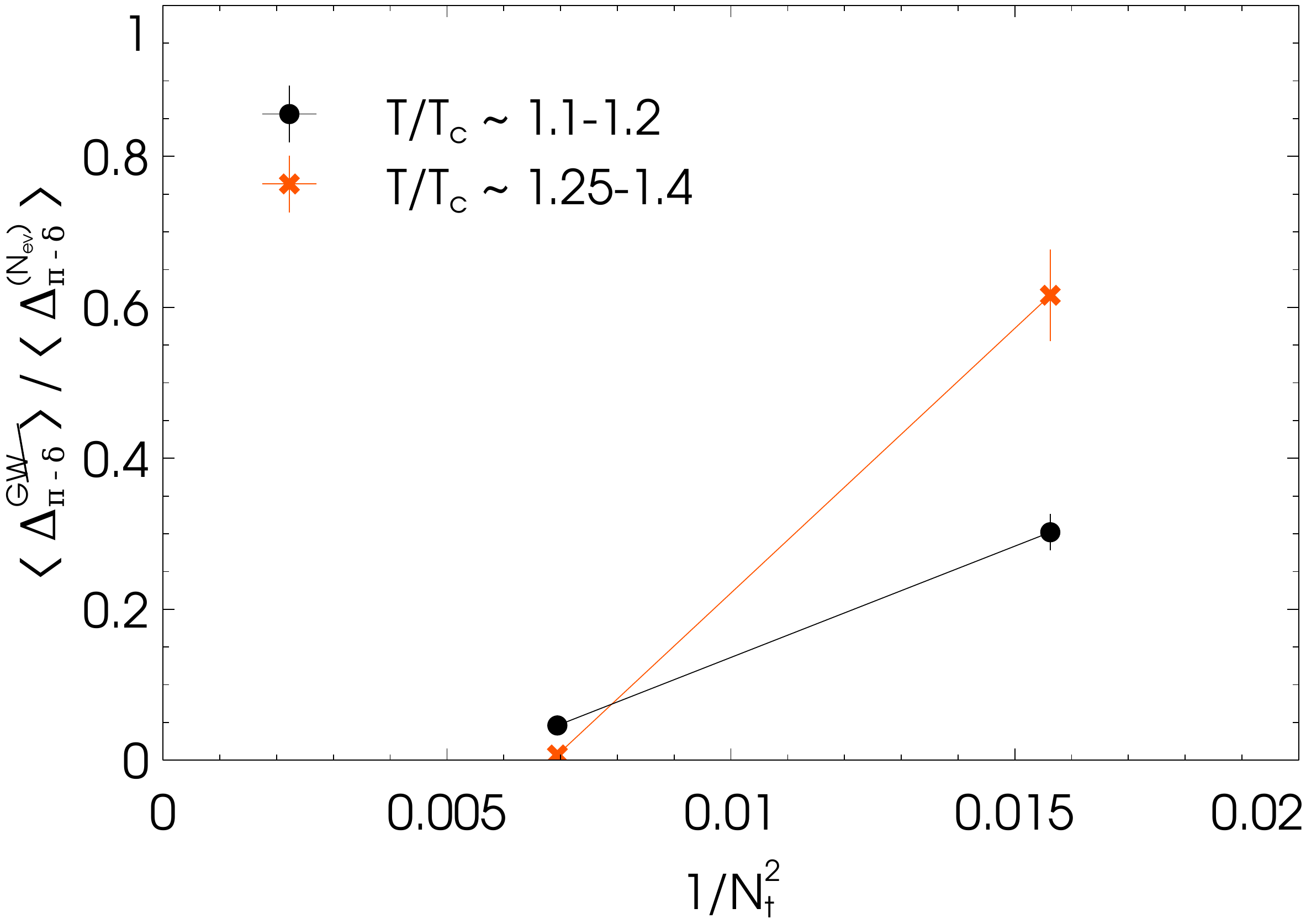}

  \caption{
    Fractional contribution of the GW violating term to
    $\langle\Delta_{\pi-\delta}\rangle$, {\it i.e.}
    $\langle\Delta_{\pi-\delta}^{\cancel{\rm GW}}\rangle /
    \langle\Delta_{\pi-\delta}^{(N_{\rm ev})}\rangle$,
    as a function of $1/N_t^2$.
    For a constant temperature, or $T/T_c$, it represents the
    dependence on $a^2$.
    We plot the lattice data corresponding to $T/T_c\simeq$ 1.1--1.2
    (circles) and to 1.25--1.4 (crosses).
    The quark mass is $m=0.01$ for all the data points which are respectively $\beta=4.10, 4.24$ and $\beta= 4.18, 4.30$. 
  }
  \label{fig:aDependence}
\end{figure}

Dependence on the lattice spacing is shown in
Figure~\ref{fig:aDependence}, where
$\langle\Delta_{\pi-\delta}^{\cancel{\rm GW}}\rangle /
 \langle\Delta_{\pi-\delta}^{(N_{\rm ev})}\rangle$
is plotted as a function of $1/N_t^2$.
For the same (or similar) temperature, it effectively shows the
dependence on the lattice spacing squared $a^2$.
The plot clearly shows that the GW violating contribution 
$\langle\Delta_{\pi-\delta}^{\cancel{\rm GW}}\rangle$
is substantial for the lattices of $N_t=8$.
It can be as large as 30\% or even 60\% of the total, which
is clearly not in the region where the usual $O(a^2)$ scaling toward
the continuum limit is applied.
Such artifact due to finite lattice spacing is significantly reduced
at $N_t=12$.

\begin{figure}[tbp]
  \includegraphics[width=.6\textwidth]{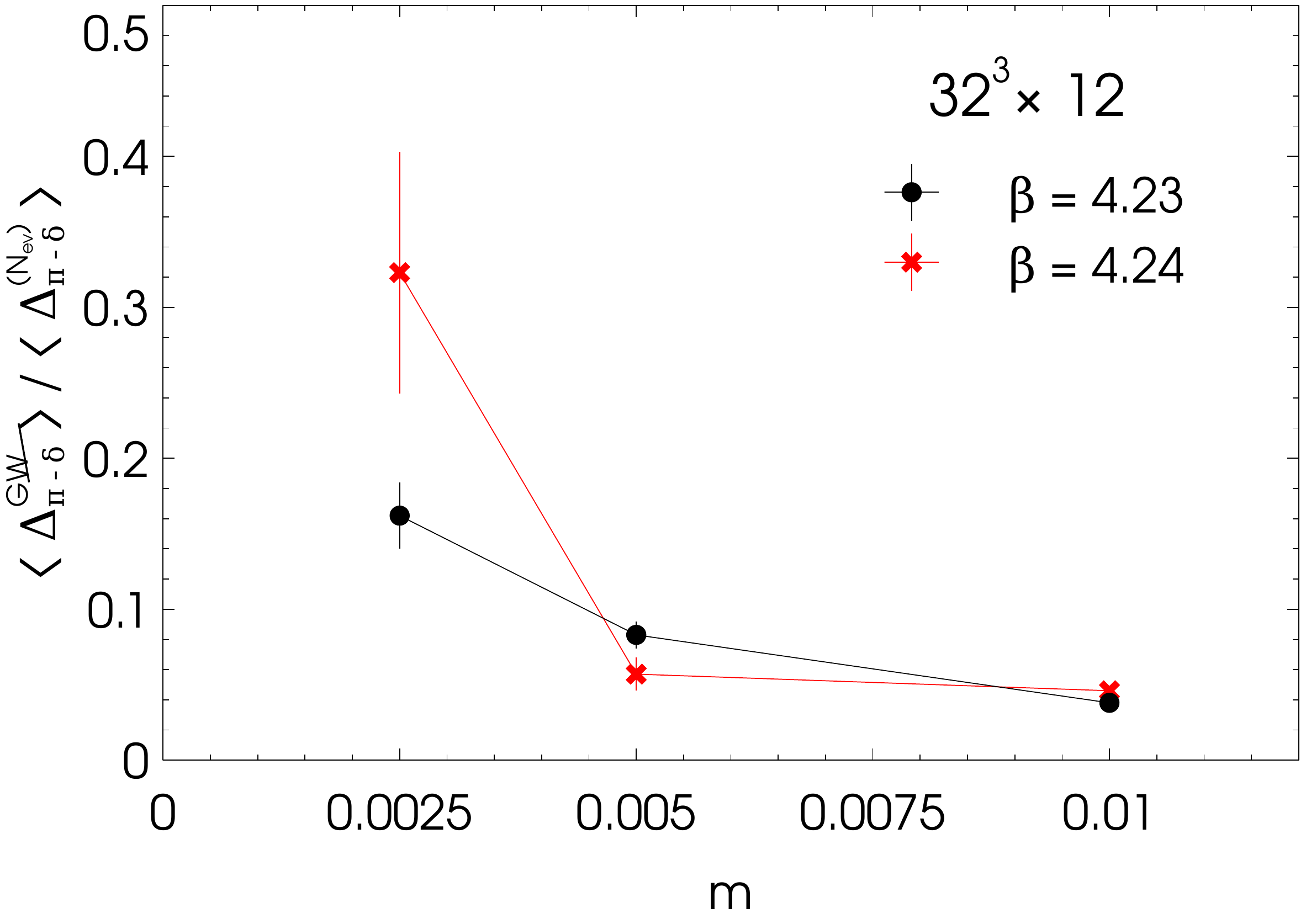}
  \caption{
    Fractional contribution of the GW violating term to
    $\langle\Delta_{\pi-\delta}\rangle$, {\it i.e.}
    $\langle\Delta_{\pi-\delta}^{\cancel{\rm GW}}\rangle /
    \langle\Delta_{\pi-\delta}^{(N_{\rm ev})}\rangle$,
    as a function of $m$.
    The lattice data at
    $\beta=4.23$, $32^3\times 12\,(\times 16)$ (circles) and
    $\beta=4.24$, $32^3\times 12\,(\times 16)$ (crosses) are plotted. 
  }
  \label{fig:MassDependence}
\end{figure}

Figure~\ref{fig:MassDependence} shows the dependence of
$\langle\Delta_{\pi-\delta}^{\cancel{\rm GW}}\rangle /
 \langle\Delta_{\pi-\delta}^{(N_{\rm ev})}\rangle$
on the quark mass $m$.
The data are those of $N_t=12$, for which the lattice artifact is
relatively small.
We observe that the GW violating term is strongly enhanced for small
quark masses. 

\begin{figure}[tbp]
  \includegraphics[width=.6\textwidth]{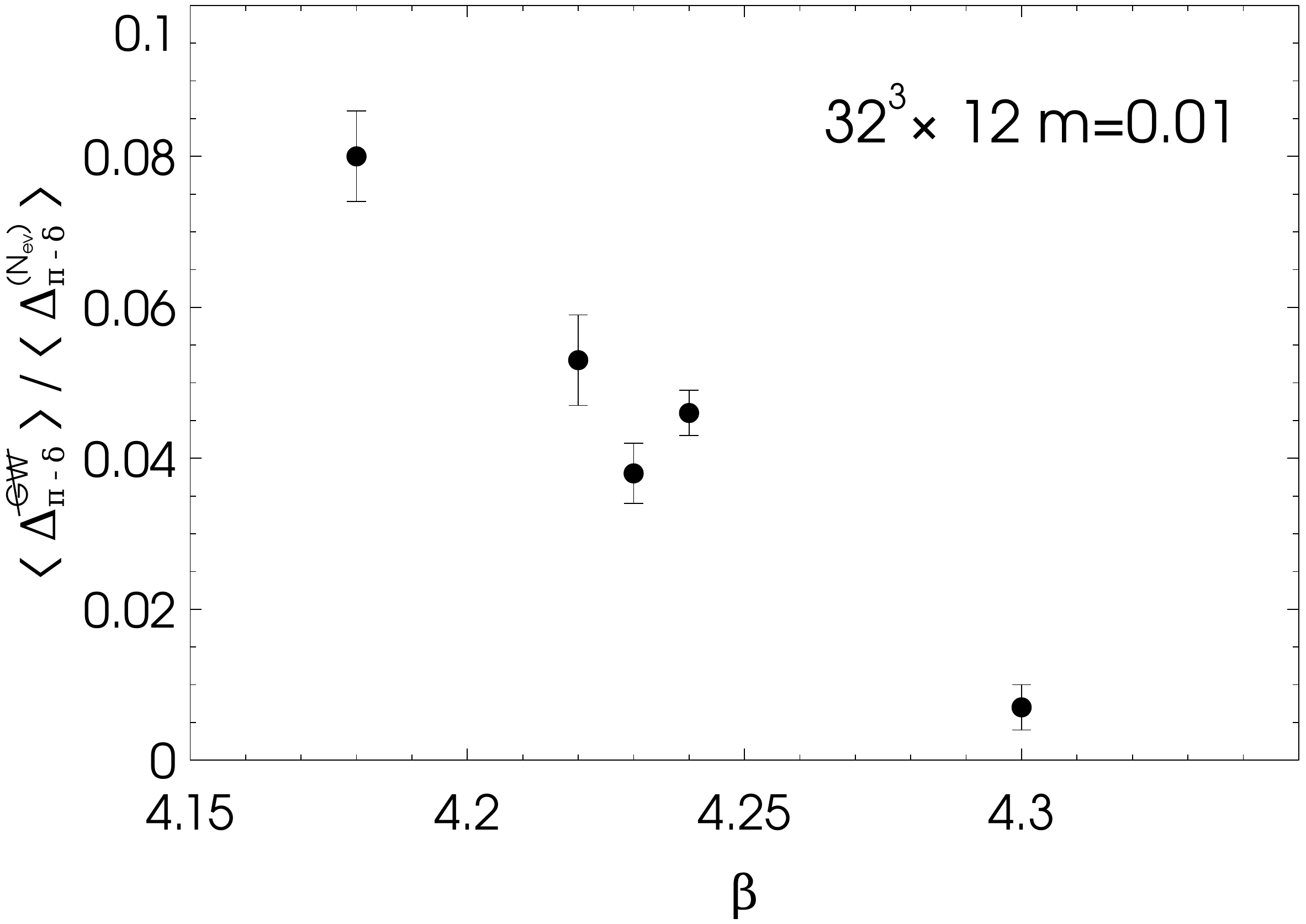}
  \caption{
    Fractional contribution of the GW violating term to
    $\langle\Delta_{\pi-\delta}\rangle$, {\it i.e.}
    $\langle\Delta_{\pi-\delta}^{\cancel{\rm GW}}\rangle /
    \langle\Delta_{\pi-\delta}^{(N_{\rm ev})}\rangle$,
    as a function of $\beta$.
    At a fixed $N_t$, this effectively represents the dependence on
    temperature. 
  }
  \label{fig:BetaDependence}
\end{figure}

By varying $\beta$ at a fixed $N_t$ (=12) we may study the
dependence of the GW violation on temperature.
From $\beta=4.18$ to 4.24 at $N_t=12$, the system goes across the
critical temperature. 
Figure~\ref{fig:BetaDependence} shows how
$\langle\Delta_{\pi-\delta}^{\cancel{\rm GW}}\rangle /
 \langle\Delta_{\pi-\delta}^{(N_{\rm ev})}\rangle$
depends on $\beta$.
We clearly see an increase of the violating term towards the lower
$\beta$ values, which can be understood partly as an effect of larger
lattice spacing.
It could be enhanced further by the accumulation of low-lying
modes in the chirally broken phase.

\subsection{Residual mass}
\label{sec:ResMass}
In this subsection we numerically study the eigenmode decomposition of
the residual mass, starting from the expression \eqref{eq:ResMassEmodes}.

\begin{figure}[tbp]
 \includegraphics[width=.48\textwidth]{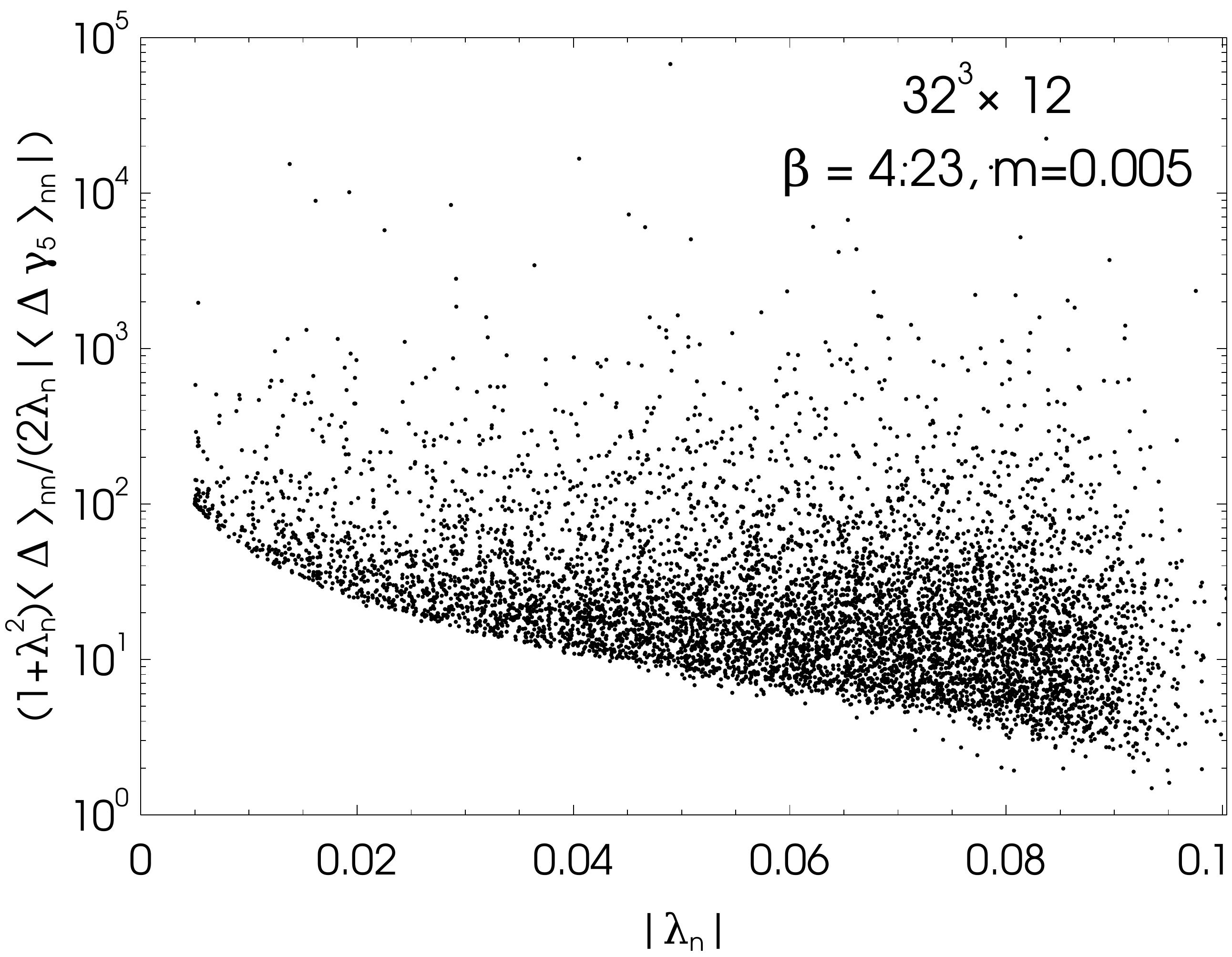}
  \caption{
    Ratio of the first term on the numerator in 
    the eigenmode decomposition of $m_{\rm res}$
    \eqref{eq:ResMassEmodes} to the second term.
    Explicitly, it is written as
    $|(1+\lambda_n^2)\langle\Delta\rangle_{nn}/
    (2\lambda_n\langle\Delta\gamma_5\rangle_{nn})|$.
    The plot shows the value for each eigenmode as a function of
    $|\lambda_n|$.
    Data at $\beta=4.23$, $32^3\times 12\,(\times 16)$, $m=0.005$ are
    plotted.
  }
  \label{fig:HGdominance}
\end{figure}

We first show that the second term on the numerator of
(\ref{eq:ResMassEmodes})
has only subdominant contributions.
In Figure~\ref{fig:HGdominance}, we plot
$|(1+\lambda_n^2)\langle\Delta\rangle_{nn}/
(2\lambda_n\langle\Delta\gamma_5\rangle_{nn})|$
for each eigenmode as a function of $|\lambda_n|$.
This corresponds to a ratio of the first term to the second in the
numerator of (\ref{eq:ResMassEmodes}).
We find that the first term is typically 10--100 larger than the
second term especially for the low-lying modes.
It implies that we can safely neglect the term of
$\langle\Delta\gamma_5\rangle_{nn}$ in the evaluation of $m_{\rm res}$.

The same observation applies also for the denominator, {\it i.e.}
the contribution of the second term
$2\langle\gamma_5\rangle_{nn}/\lambda_n$ is negligible compared to the
first $(1+\lambda_n^2)/\lambda_n^2$.
The residual mass is then precisely approximated by a weighted average
of $\langle \Delta \rangle_{nn}$ of the form
\begin{equation}
  m_{\rm res} \simeq 
  \frac{
    \displaystyle
    \sum_n\frac{1+\lambda_n^2}{\lambda_n^2} \langle\Delta\rangle_{nn} 
  }{
    \displaystyle
    \sum_n \frac{1+\lambda_n^2}{\lambda_n^2}
  }.
\end{equation}
We can therefore gain a rough idea of the residual mass by inspecting
$\langle\Delta\rangle_{nn}$. 
From the plots in Figure~\ref{fig:DeltaBin} we find that 
$\langle\Delta\rangle_{nn}$ is approximately constant in the lowest
part of the spectrum. 
On the other hand, we can calculate $m_{\rm res}$ using a stochastic
method, the result of which is shown by a dot-dashed line in the plots
of Figure~\ref{fig:DeltaBin}. 
We find that the the contribution from the individual low-lying
eigenmodes $\langle\Delta\rangle_{nn}$ is typically one order of
magnitude larger than the weighted average $m_{\it res}$, while the
average of $\langle\Delta\rangle_{nn}$ in a bin of $|\lambda_n|$ tends
to decrease towards the stochastic estimates for larger $|\lambda_n|$.
It implies that the GW violating effect is enhanced in the low-lying
modes.

\begin{table}[tbp]
  \begin{center}
    \begin{tabular}{ccccc}
      \hline
      $\beta$ & $m$ & $N_s^3\times N_t(\times L_s)$ 
      &$m_{\rm res}$ ($\lambda_n < 0.08$) & $m_{\rm res}$ (stochastic)
      \\
      \hline 
      4.10& 0.01  & $32^3\times 8\,(\times 12)$ & 0.0065(5) & 0.0010(2)\\
      4.10& 0.005 & $32^3\times 8\,(\times 24)$ & 0.0030(3) & 0.00053(4)\\
      4.18& 0.01  & $32^3\times 12\,(\times 16)$ & 0.0010(1) & 0.00022(2)\\
      4.23& 0.01  & $32^3\times 12\,(\times 16)$ & 0.00047(5)& 0.00010(1)\\
      4.24& 0.01  & $32^3\times 12\,(\times 16)$ & 0.00068(12)& 0.00009(1)\\
      4.23& 0.005 & $32^3\times 12\,(\times 16)$ & 0.00063(6)& 0.00012(2)\\
      4.24& 0.005 & $32^3\times 12\,(\times 16)$ & 0.00048(7)& 0.00010(2)\\
      4.23& 0.0025& $32^3\times 12\,(\times 16)$ & 0.00066(7)& 0.00016(4)\\
      4.24& 0.0025& $32^3\times 12\,(\times 16)$ & 0.00107(17)&0.00013(3)\\
      \hline
    \end{tabular}
    \caption{
      Residual mass calculated using the lowest part of the spectrum
      ($\lambda_n < 0.08$) compared with full result obtained with a
      stochastic measurement.
    }
    \label{table:ResMass}
  \end{center}
\end{table}

We estimated the residual mass in the low-mode region by applying an
arbitrary upper-cut, that we chose $\lambda_\text{cut} = 0.08$, to the
eigenvalue sums in the numerator and in the denominator.
The results are listed in Table~\ref{table:ResMass}.
They confirm that in this region of the spectrum the effective
residual mass is significantly larger, as expected from the
observation of $\langle\Delta\rangle_{nn}$. 

The residual mass as obtained with \eqref{eq:ResMassEmodes} is 
dominated by the ultraviolet part of the eigenmodes because of the
increasing number of the eigenmodes $\sim\lambda^3$, even though the
weight factor $\frac{1+\lambda_n^2}{\lambda_n^2}$ would favor the
infrared region.
This is qualitatively in agreement with the fact that the residual
mass calculated with the pion external state (\ref{eq:resmass_pi}) is
significantly larger than the full summation of the space-time points (\ref{eq:resmass_full}).
(See a discussion in \cite{Boyle:2015rka} for more details). 


\section{Summary}
The remnant violation of the GW relation is a potential source of
substantial systematic error for some physical quantities, for which
the low-lying eigenmode give a significant (or even dominant) contribution.
In this work we have shown how to identify the effect of the violation
in the meson susceptibilities, chiral condensate and the residual mass
by decomposing their matrix elements in the Dirac eigenmode basis. 
We obtained exact equations that account for the violation terms and
allow their quantitative estimate. 
All the violation terms can be described in terms of the matrix elements of $\gamma_5$ in the eigenmode basis. 

Numerical calculations show that the difference of susceptibilities
$\Delta_{\pi -\delta}$ can be strongly affected by the violations
coming from the lowest part of the spectrum on coarse lattices. 
The signal for M\"obius domain-wall fermions at lattice spacings 
$>0.1$~fm is dominated by lattice artifacts making the naive
calculation unreliable. 
At finer lattices, $\sim 0.08$~fm, the effect is reduced dramatically
to a more manageable level, 10--20\%. 
The mass dependence, showing an increase of the effect in the
$m\rightarrow 0$ limit, confirms that the lowest-eigenmode region is
critical and finer lattices are necessary for a proper chiral limit
using the M\"obius domain-wall fermions.

The analogous expansion of the residual mass shows another example
that the lowest part of the spectrum has larger violations to the
Ginsparg-Wilson relation. 
The estimate of $m_\text{res}$ in this spectral region shows that the
naive estimate is about one order of magnitude smaller, potentially
underestimating the GW-violating contributions.
The effect of the remnant chiral symmetry violation needs to be
estimated for individual quantities of interest, and the definition of
$m_{\rm res}$ constructed from the low-lying modes is more adequate for the
quantities dominated by the lowmodes.

\begin{acknowledgments}
Numerical simulations are performed on IBM System Blue Gene Solution at High Energy Accelerator Research Organization (KEK) under a support for is Large Scale Simulation Program (No. 14/15-10). This work is supported in part by the Grant-in-Aid of the Ministry of Education (No. 25800147, 26247043, 15K05065) and by MEXT SPIRE and JICFuS. 
\end{acknowledgments}

\newpage
\bibliography{references}

\end{document}